\documentclass[12pt,preprint]{aastex}

\newcommand{\eg}{{\it e.g.}}
\newcommand{\etal}{et~al.}
\newcommand{\ic}{$I_{\rm C}$}

\newcommand{\mdot}{$\dot{M}$}
\newcommand{\msun}{M$_{\sun}$}

\newcommand{\vi}{$(V-I_{\rm C})$}
\newcommand{\vsini}{$v \sin i$}
\newcommand{\mum}{$\mu m$}

\received{}
\revised{}
\accepted{}
\shorttitle{}
\shortauthors{Rebull \etal}

\begin{document}

\title{A Correlation Between Pre-Main Sequence Stellar Rotation
Rates and IRAC Excesses in Orion}
\author{L.\ M.\ Rebull\altaffilmark{1}, J.\ R.\ Stauffer\altaffilmark{1},   
        S.\ T.\ Megeath\altaffilmark{2}, J.\ L.\ Hora\altaffilmark{2}, 
	L.\ Hartmann\altaffilmark{2,3}}

\altaffiltext{1}{Spitzer Science Center, Caltech M/S 220-6, 1200 E.\
California Blvd., Pasadena, CA 91125 (luisa.rebull@jpl.nasa.gov)}
\altaffiltext{2}{Harvard-Smithsonian Center for Astrophysics}
\altaffiltext{3}{University of Michigan-Ann Arbor}

\begin{abstract}

Early observations of T~Tauri stars suggested that stars with
evidence of circumstellar accretion disks rotated slower than
stars without such evidence, but more recent results are not as
clear. Near-IR circumstellar disk indicators, though the most
widely available, are subject to uncertainties that can result
from inner disk holes and/or the system inclination. 
Mid-infrared observations are less sensitive to such effects,
but until now, these observations have been difficult to
obtain.  The Spitzer Space Telescope now easily enables
mid-infrared measurements of large samples of PMS stars covering
a broad mass range in nearby star-forming regions. Megeath and
collaborators surveyed the Orion Molecular Clouds ($\sim$1 Myr)
with the IRAC instrument  (3.6, 4.5, 5.8, 8 \mum) as part of a
joint IRAC and MIPS GTO program.  We examine the relationship
between rotation and Spitzer mid-IR fluxes for $\sim$900 stars
in Orion for stars between 3 and 0.1 \msun.  We find in these
Spitzer data the clearest indication to date that stars with
longer periods are more likely than those with short  periods to
have IR excesses suggestive of disks.

\end{abstract}


\section{Introduction}
\label{sect:intro}

The smallest molecular cores observed to date have at least
$\sim$6 orders of magnitude greater angular momentum per unit
mass than the Sun, suggesting that they would eventually greatly
exceed the breakup velocity if no angular momentum was lost
during the evolutionary process.  It is widely believed that
most if not all low-mass stars form with circumstellar accretion
disks; accreting solar-like pre-main sequence (PMS) stars have
rotational velocities 1/5th to 1/10th of breakup speeds despite
the angular momentum transported to the star via accretion from
its circumstellar disk.  These fundamental observations lead to
the conclusion that an angular momentum regulation mechanism
must be at work during the accretion phase.  Stellar winds alone
as conventionally invoked cannot explain such angular momentum
loss unless they operate completely differently than expected
(see, e.g., Sills \etal\ 2000 and references therein); these
young solar-like PMS stars are completely convective, so winds
must slow the rotation of the whole star (not just the outer
layers), and the timescale for winds to significantly slow a
star's rotation is longer than the evolutionary timescales
involved here.  

All recent thinking focuses on angular momentum transfer
mechanisms intimately related to the accretion process itself.
These include (a) transfer of stellar angular momentum to a
spreading accretion disk (\eg\ K\"onigl 1991, K\"onigl \&
Pudritz 2000), and (b) transfer of angular momentum to an
accretion-driven wind, launched either at the disk/magnetosphere
boundary (\eg\ Shu \etal\ 2000) or elsewhere in the
star-accretion disk system (Matt \& Pudritz 2005). All of the
above disk-related regulation mechanisms posit that an accretion
disk is present, and that the central star is `locked' (or
nearly so) to a constant angular velocity fixed by the Keplerian
speed at the radius at which the stellar magnetosphere and disk
are linked (co-rotation radius) even as the PMS stars contract.

For more than 10 years (including, e.g., Edwards \etal\ 1993),
observers have attempted to test the hypothesis that
disk-related processes account for angular momentum regulation
among solar-like PMS stars. The test is seemingly simple:
compare the observed rotation periods or projected equatorial
rotational velocities of PMS stars surrounded by disks (and
thus, by hypothesis, regulated) with those that lack disks and
are presumed free to spin up in response to contraction toward
the main sequence absent a regulating disk. The results of these
tests have, however, been mixed; for example, Rebull (2001) and
Makidon \etal\ (2004) find no clear correlation between disk
excess and rotation rate in Orion and NGC 2264, respectively,
but Herbst \etal\ (2002) and Lamm \etal\ (2004) find such a
correlation in these same clusters.  Moreover, Herbst \etal\
(2002) find a weak correlation between the rotation rate and the
{\em size} of the disk excess, which is not found by, e.g.,
Stassun \etal\ (1999) or Rebull (2001).

Two key factors complicate this apparently simple test and may
lead to ambiguous results: (a) the inaccuracies inherent in
near-IR excesses, and (b) the influence of sample size.  Near-IR
excesses (typically $K$-band), although the most widely
available, are not sufficiently robust, owing to the presence of
inner disk holes and projection effects (see, e.g., Hillenbrand
et al.\ 1998) and/or contrast between the disk and the
photosphere and/or disk geometry (e.g., the size of the inner
disk wall).  These young stars are variable, and use of near-IR
indicators is often also compromised by the vexing problem of
non-simultaneity in the photometric observations (optical
through infrared) typically used to derive $K$-band excesses.  
A recent paper by Rebull \etal\ (2004) identifies the second key
factor as the sample size.  Rotation periods among young
PMS stars surrounded by disks span a factor of 20.  The broad
period distribution among young PMS stars means that extracting
a `signal' of period changes between samples of stars surrounded
by disks and those that lack disks requires a large sample
before it is possible to distinguish `locked' from `freely
spinning' objects.  The fact that near-IR excesses `miss'
$\sim$30\% of stars which longer wavelength observation reveal
to have clear disk signatures complicates separation of disked
and non-disked samples -- a potentially fatal difficulty for
small samples ($<$ several hundred) stars.

We now are in a position to overcome both of these problems.
Over the past decade, more than 1700 rotation periods have been
measured for solar-like PMS stars, $\sim$900 in Orion alone.
Moreover, the advent of the Spitzer Space Telescope (Werner
\etal\ 2004) provides a powerful tool for robust identification
of stars surrounded by disks, via observation of the IRAC
channel 1 (3.6 \mum) $-$ IRAC channel 4 (8 \mum) color index --
an index both highly diagnostic of emission arising in
circumstellar accretion disks, and little influenced by
inclination effects, inner disk holes or (because of the method
of data acquisition) non-simultaneity of photometric
observations.

In this contribution, we discuss the period distributions for a
sample of $\sim$500 stars drawn from the Orion region ($\sim$1
Myr) for stars with and without circumstellar disks as diagnosed
by [3.6 \mum] $-$ [8 \mum] colors (hereafter denoted
$[3.6]-[8]$). To anticipate our conclusions, we find that only a
small fraction of rapidly rotating stars exhibit robust disk
signatures, while the fraction of disks surrounding slowly
rotating stars is much greater (with a high degree of
statistical significance).  We find in these Spitzer data the
clearest indication to date that stars with longer periods are
more likely than those with short periods to have IR excesses
suggestive of disks.

\section{Data }
\label{sec:data}

\subsection{Sources of data}

\begin{figure*}[tbp]
\epsscale{0.5}
\plotone{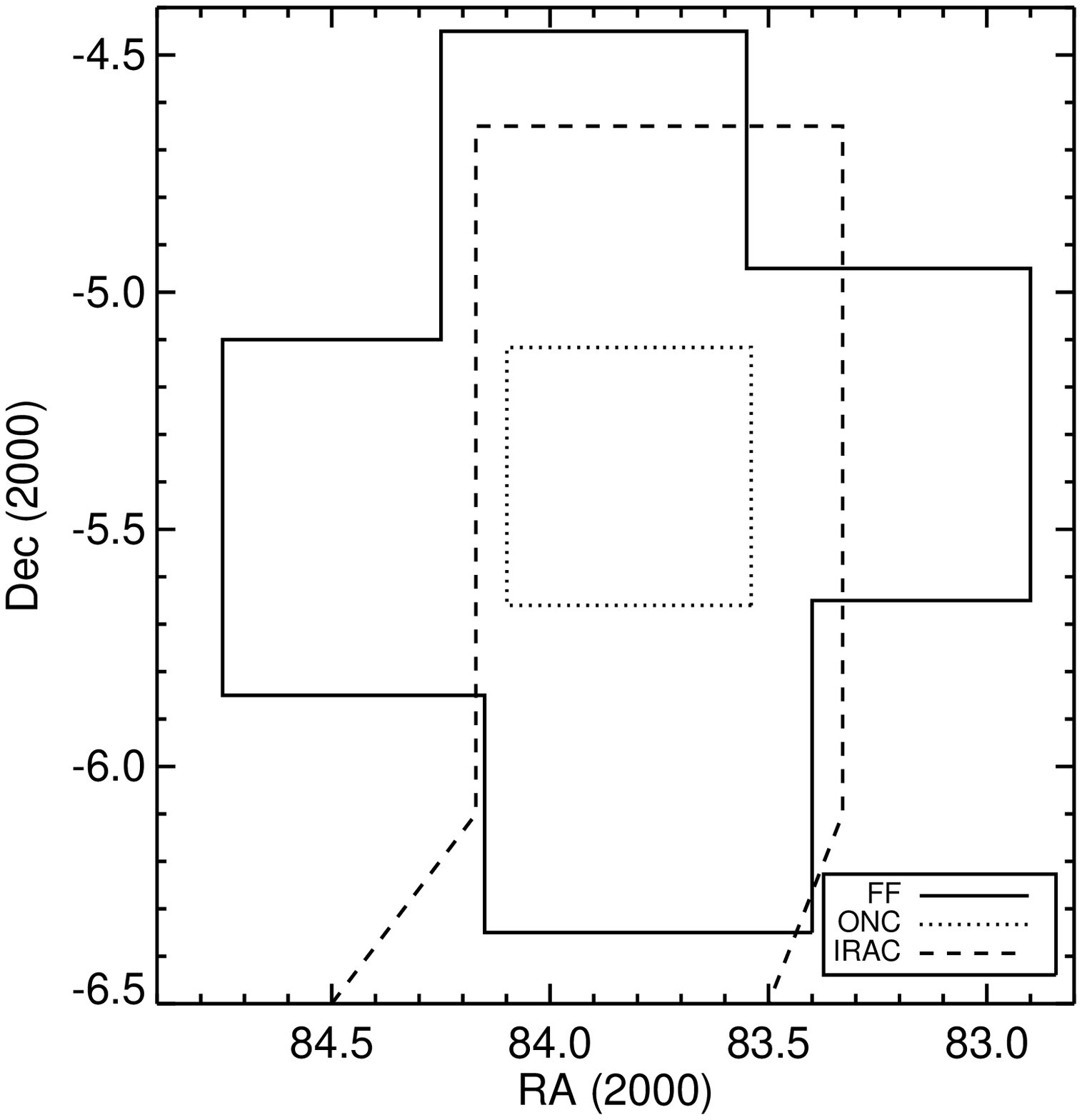}
\caption{A map of the region covered by the IRAC study relevant
to the present discussion of rotation rates. The central box
(dotted line) is the `traditional' ONC region, the four square
fields (solid line) to the N, S, E, and W of the central region
are the Flanking Fields (FF), and the dashed lines are the
extent of the IRAC survey, which continues to the southeast.}  
\label{fig:where}  
\end{figure*}

Megeath et al.\ (2006, in prep) surveyed the Orion Molecular
Clouds ($\sim$1 Myr) with the IRAC instrument  (3.6, 4.5, 5.8,
\& 8 \mum; Fazio \etal\ 2004) as part of a larger joint IRAC and
MIPS GTO investigation.  The complete IRAC survey covers 6.8
square degrees (4.4 square degrees in Orion A and 2.4 square
degrees in Orion B), including the region near the Trapezium (see
Figure~\ref{fig:where}) but also continuing along the molecular
cloud to the southeast. The ``traditional'' Orion Nebula Cluster
(ONC, $\sim$0.5 Myr) is a $\sim0.5\arcdeg\times0.5\arcdeg$
region (see Fig.~\ref{fig:where}), including stars with periods
found by Herbst \etal\ (2000, 2002), Stassun \etal\ (1999), and
Carpenter \etal\ (2001). The Flanking Fields (FF; $\sim$1 Myr;
Rebull 2001 and Rebull \etal\ 2000) are four fields to the N, S,
E, and W of the ONC, each of which is $\sim$0.75$\arcdeg$ on a
side (see Fig.~\ref{fig:where}).  Stars in the FF are slightly
older than those in the ONC (Ramirez \etal\ 2004).  Periods for
stars here come from Rebull (2001), Stassun \etal\ (1999) and
Carpenter \etal\ (2001). The IRAC survey follows the molecular
cloud, even through the ONC and FF region, so it includes most
of Field 2, to the North of the ONC, and all of Field 4, to the
South of the ONC.  Less than a third of fields 1 and 3 are
included in the IRAC survey.  

See Megeath \etal\ (2006, in prep) for more extensive details on
data acquisition and reduction for the Orion data.  In summary,
toward each region, four 12-second high-dynamic-range (HDR)
frames were obtained.  The individual basic calibrated data
(BCD) frames (Ver.\ S11) were mosaicked together using the MOPEX
software.  The  sources were identified with a custom source
finder written in IDL and then measured  with the IDLPHOT
package.  An  aperture of 2.4$\arcsec$ and a sky radius of
2.4$\arcsec$ to 7.2$\arcsec$ were used.  The adopted zero
points  in units of  DN/sec/pixel are 19.66, 18.94, 16.88 and
17.39 for a 10 pixel aperture radius in IRAC channels 1 through
4 respectively.   Aperture corrections of 1.213,1.234,1.379 and
1.584 were applied to the data. Table~\ref{tab:data} contains 
IRAC fluxes for those stars with periods detected in both IRAC-1
and IRAC-4 (see below).

\subsection{Rotation rates and disk indicators}

For stars in the region covered by IRAC observations, there are
854 stars with known periods. The catalog of stars with periods
were  merged with a catalog of IRAC sources identified by an
automated  source finder developed for this program (Megeath
\etal\ 2006).  In IRAC-1 (3.6 \mum), 90\% of those stars are
detected, 96\% are detected in IRAC-2 (4.5 \mum), 81\% are
detected in IRAC-3 (5.8 \mum), and 59\% are detected in IRAC-4
(8 \mum).  (We note for completeness that the MIPS 24 \mum\
counterpart to this GTO survey retrieves only 13\% of the stars
with known periods; for this reason, the MIPS data are not
discussed here.)  Because the measurement of periods necessarily
implies optical (or NIR) detection of the stellar photosphere,
by examining only stars with measured periods, we have already
effectively selected just the less-embedded stars that are
brighter in IRAC wavelengths.  The entire IRAC survey
completeness limit in this region ranges from 3 to 7 magnitudes
deeper than that of the periodic sample, so without reference to
backgrounds, all stars with periods should be detectable in the
IRAC survey.  However, in reality the nebular background and
source confusion increases closer to the Trapezium.  The
majority of periodic stars are detected, but those periodic
stars missing IRAC detections come from the region closest to
the Trapezium, making the detection and identification of even
some bright sources less reliable.  The nebular background
increases with wavelength, so a larger fraction of stars are
missing 8 \mum\ detections than 3.6 \mum\ detections.  Since the
ONC is slightly younger than the surrounding regions (Ramirez
\etal\ 2004), this effectively means that the sample of stars
with periods and detections at the longer wavelengths will be
slightly skewed to older stars.  

Color-color plots like those appearing in Megeath \etal\ (2006,
in prep) for this sample show clear populations of disk
candidates and non-disk candidates; see, e.g., Allen \etal\
(2004) for more discussion on interpretation of IRAC colors.  We
considered several different possible disk indicators to use
here, but the separation between the disk candidates and
non-disked candidates was most apparent when using a larger
difference in wavelength.  Despite having fewer stars to
consider when using the 8 \mum\ detections, the conclusions of
this analysis do not change significantly when other disk
indicators are considered.  Therefore, we will use $[3.6]-[8]$ 
measurements to identify disks, where we have set (based on
histograms of excesses; see below) the boundary between disk and
non-disk candidates at $[3.6]-[8]$=1 magnitude.  Requiring both
3.6 and 8 \mum\ detections reduces the 854 available periods
down to 464 stars, where the stars are preferentially lost from
the core of the ONC (where the nebular emission and source
confusion are both high). These 464 stars appear in
Table~\ref{tab:data}.   None of our conclusions change
significantly when considering other disk indicators involving
at least one IRAC band (such as the longer lever arm accorded by
$K-[8]$, or the additional stars provided by $[3.6]-[5.8]$).

%We looked for but did not find any detectable trend of detection
%with color. 

In the past, it has been assumed that the distribution of
measured periods is representative of the larger distribution of
rotation rates.  Samples of stars with and without measured
periods have similar \vsini\ distributions (e.g., Rhode \etal\
2001).  However, recently, the sample of stars with measured
periods has been shown to be significantly brighter in X-rays
than the sample of stars without measured periods (e.g., Rebull
\etal\ 2006 in press), suggesting that there may be real
differences in the two samples.  We looked for any difference in
IRAC properties between those stars with measured periods and
those stars without measured periods.  In order to assemble a
fair comparison, we only considered objects (either with
measured periods or without) also detected in \ic\ (values taken
from the literature) and those located in a clear ``locus''
above the zero-age main sequence (ZAMS) in the \ic/\vi\
color-magnitude diagram (see Rebull \etal\ 2000 for much more
discussion, including CMDs). 

\begin{figure*}[tbp]
\epsscale{0.5}
\plotone{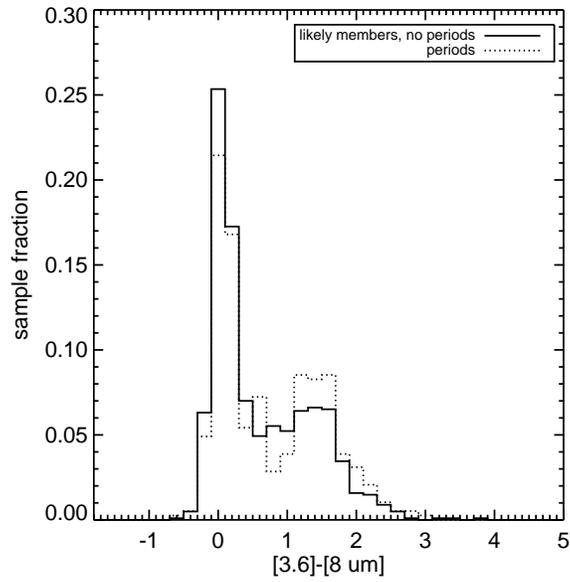}
\caption{ Normalized histograms of $[3.6]-[8]$ for stars with
and without periods that are also detected in \ic\ and likely
cluster members based on position in a color-magnitude diagram
(see text), for stars throughout the Orion region.  The
distributions are statistically indistinguishable. }  
\label{fig:all14_histo}  
\end{figure*}

Figure~\ref{fig:all14_histo} shows normalized histograms of the
$[3.6]-[8]$ colors for stars with measured periods and those
otherwise similar stars that are likely cluster members but
without measured periods.  There is a clear separation of the
disk candidates above $[3.6]-[8]$=1 and the non-disk candidates
below that level.  This demonstrates vividly the advantage of
using MIR disk indicators; this effect is not found so
prominently using other (e.g., NIR) disk indicators, which
indeed has complicated earlier searches for disks.  We note that
the separation of the disk candidates at $[3.6]-[8]$=1 is
slightly clearer for the stars with periods than the stars
without measured periods; we suspect that this is because the
sample of stars without measured periods are more likely to have
non-cluster member interlopers.  We have normalized the
histograms in the Figure because of the vastly different numbers
of stars (1014 without periods and 387 with periods). A
Kolmogorov-Smirnoff (K-S) test on these (un-normalized)
distributions reveal that they are statistically
indistinguishable, and we conclude that the IRAC disk properties
of the stars with and without measured periods are statistically
indistinguishable.

\section{Correlation with Rotation}

\begin{figure}
\plotone{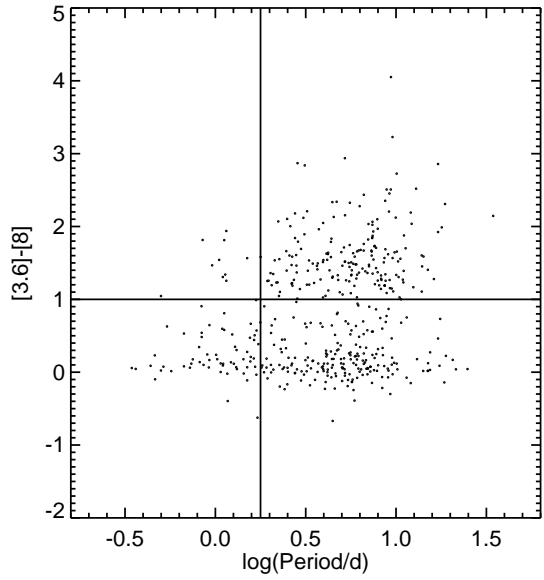}
\caption{ Plot of $[3.6]-[8]$ vs.\ period for stars throughout
the Orion region. There is a clear separation of the disk
candidates above (and the non-disk candidates below)
$[3.6]-[8]$=1, and the distribution of excesses is clearly
different  for periods $>$ and $<$ log $P$=0.25; see next
figure.}  
\label{fig:p14}  
\end{figure}

Figure~\ref{fig:p14} plots $[3.6]-[8]$ vs.\ period for all stars
throughout the Orion region.  There is still a clear separation
of the disk candidates above $[3.6]-[8]$=1 and the non-disk
candidates below that level.   By examining the cumulative
distribution function of all $[3.6]-[8]$ colors, we find an
inflection in the curve near $[3.6]-[8]$=1, verifying what we
see by eye in the distribution of colors in Fig.~\ref{fig:p14}
(or Fig.~\ref{fig:all14_histo}).

\begin{figure}
\plotone{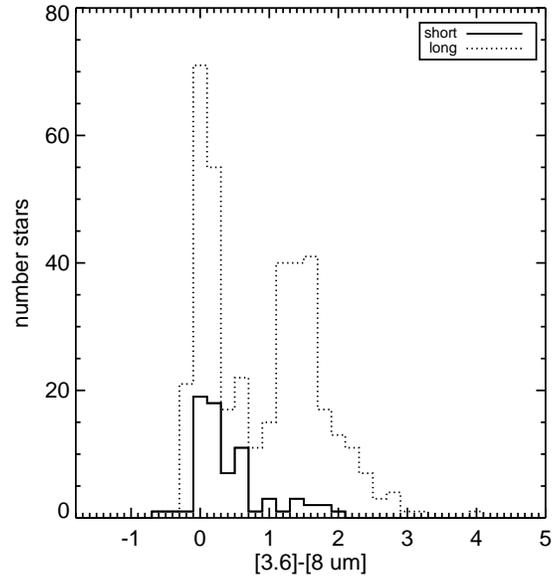}
\caption{ Histograms of $[3.6]-[8]$ for stars with rotation
rates $<$ and $>$ log $P$=0.25 ($P$=1.8d), for stars throughout
the Orion region.  The distributions are significantly different;
see next figure.}  
\label{fig:p14_histos}  
\end{figure}

It is similarly already apparent that the distribution of
$[3.6]-[8]$ is different for shorter than it is for longer
periods.   By examining the cumulative distribution function of
periods for all stars with  $[3.6]-[8]>1$, there is an
inflection point near $P\sim$1.8 d (log $P\sim$0.25). 
Figure~\ref{fig:p14_histos} shows the histograms of $[3.6]-[8]$,
distinguishing  $P<$1.8 d (log $P<$0.25), and $P>$1.8 d.
Clearly, a long-period star is more likely to have a disk than a
short-period star, but there are also substantial numbers of
long-period stars with little or no excess.  These latter stars
may have just recently cleared their disks and have not yet spun
up in response to contraction on their way to the ZAMS.

\begin{figure}
\plotone{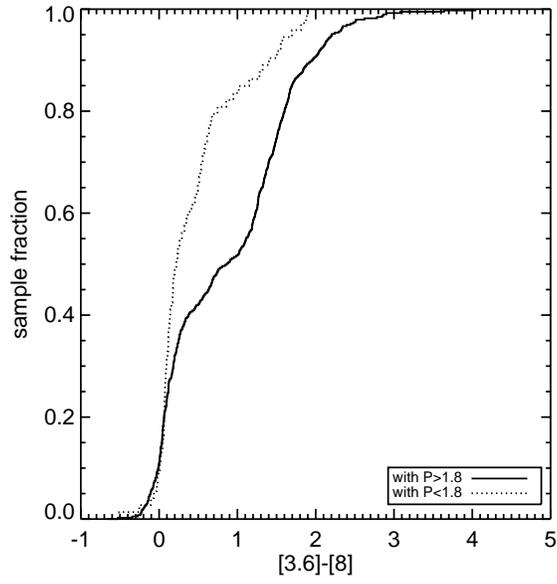}
\caption{Cumulative distributions of $[3.6]-[8]$ for stars
with rotation rates $<$ and $>$ log $P$=0.25 ($P$=1.8d), for stars
throughout the Orion region.  The distributions are significantly
different.}  
\label{fig:cumudist_14}  
\end{figure}

Figure~\ref{fig:cumudist_14} shows the cumulative distributions
of [3.6]$-$[8] for the samples with $P>$ and $<$1.8 d portrayed
in Fig.~\ref{fig:p14}.  (The inflection point mentioned above
near $[3.6]-[8]=1$ is readily apparent.)  A K-S test on these
distributions reveal that they are statistically substantially
different ($<$1 chance in $10^6$ that they are drawn from the
same population). The distributions of $P$ for samples with
[3.6]$-$[8]$>$ and $<$1 are also statistically different. 

There are 11 fast-rotating stars that appear to have disks.  Of
these 11 stars, one (JW 783) is a likely double in the IRAC
images, and one (Par 2097) falls in a particularly bright region
of the nebula (and therefore is likely to have more inaccurate 8
\mum\ fluxes than typical). The remaining 9 stars probably have
reasonable measurements at 3.6 and 8 \mum, despite many of them
having some nebular background at 8 \mum. Perhaps these fast
rotating stars with disks have just recently experienced an
accretion event, or they are descendents of EX Ori-type stars.
These 9 stars are good targets for followup studies, and they
are Par 1409, Par 1799, Par 1938, Par 2123, JW 104, JW 879 (a
likely non-member based on proper motions), numbers 1053 and
3115 from Hillenbrand (1997), and number 864 from Rebull (2001).

% foo1   good        hint of artifact: prob.  OK
% foo2    good       faint star, mod  nebula: cont.
% foo3    crowded, near bright source:  prob  OK
 % foo4    good       faint star, bright  nebula:  bad
% foo5    good       good
% foo6    good       moderate 8 um  nebulosity: some cont.
 % foo7    double
% foo8    good       moderate 8 um  nebulosity: some cont
% foo9    good       faint 8 um nebula  :   OK
% foo10  good       moderate 8 um  nebulosity: come cont
% foo11 good        faint star,  nebulosity cont
%
%foo1       83.577354      -5.0799830 Par 1409
%foo2       83.798450      -5.2826970 Par 1799
%foo3       83.833532      -5.3516470 Par 1938
%foo4       83.885503      -5.4362360 Par 2097
%foo5       83.901841      -5.5697630 Par 2123
%foo6  83.704489d      -5.4407280d  jw 104 
	% this one has log p=0.25042
%foo7       83.856158      -5.5680082 JW 783
%foo8       83.883176      -5.2724908 JW 879
%foo9       83.633420      -5.4618410 H97: 3115
%foo10       84.016086      -5.5052250 H97: 1053
%foo11       83.491305      -5.6074960 R01- 864

We expected that there might be differences in this relation
between $P$ and IR excess as a function of location within the
cluster (ONC vs.\ FF).  There are differences between the ONC
and the FF, but the differences are all in the relative disk
fractions.  The disk fraction is much higher in the ONC, and
although this could be a result of the fact that the ONC is
slightly younger than the FF (Ramirez \etal\ 2004), this is
likely also a result of the bias against faint 8 \mum\
detections of stars without bright infrared excesses in the ONC
imposed by the weaker photospheric emission and substantially
higher background seen at longer wavelengths.  As a function of
physical location, there is no distinguishable difference of the
clear separation of disk candidates from non-disk candidates, or
of the tendency for fast rotators to lack disks.

We expected that there might be an effect of mass or age.  We
derived masses and ages via placement in the $I/V-I$
color-magnitude diagram and via D'Antona and Mazzitelli
(1994,1998; DAM) models, as in Rebull (2001), with conversion
from the theoretical to the observational plane using
transformations found in Hillenbrand (1997).  (We note that of
course masses derived in this fashion are model-dependent. We
could have investigated mass effects via the use of spectral
types, but types only exist for 50\% of our sample.  In lieu of
cutting the sample down by half, we have opted instead to
proceed using masses so derived.) Previous investigators (e.g.,
Herbst \etal\ 2002) report different period distributions for
stars above and below a mass division of 0.25 \msun, also using
DAM models. Therefore, we also investigated separating the
sample at this boundary, corresponding roughly to a spectral
type of M2.5-M3, or a mass of $\sim$0.35 \msun\ using Siess
\etal\ (2000) models (the Z=0.02 model with no overshooting,
using conversions from Kenyon \& Hartmann 1995), or roughly
$\sim$0.5 \msun\ using Baraffe \etal\ (1998) models (Z=0, He
mass fraction=0.275, mixing length=1.0, conversions as presented
therein).

\begin{figure}
\plottwo{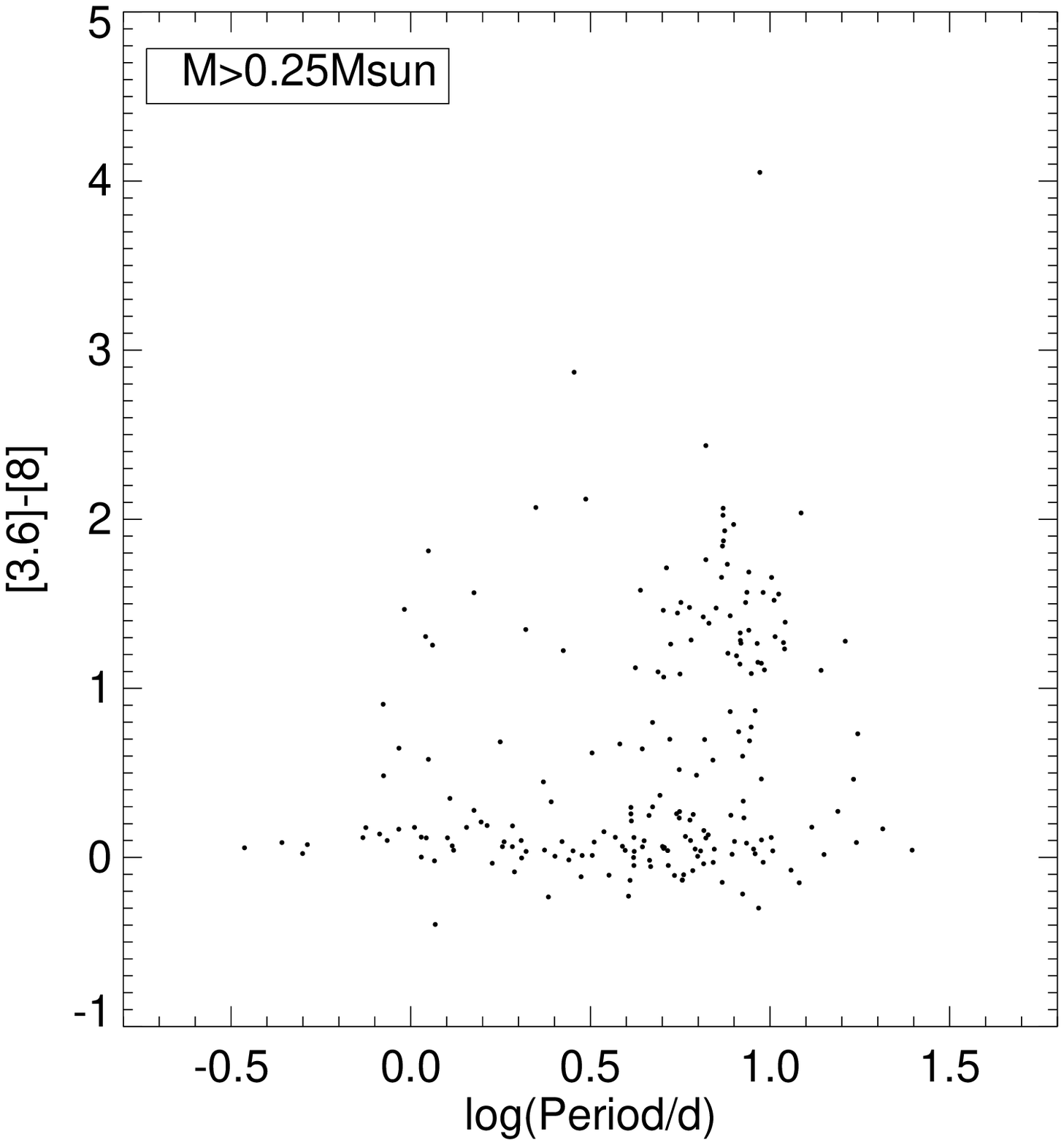}{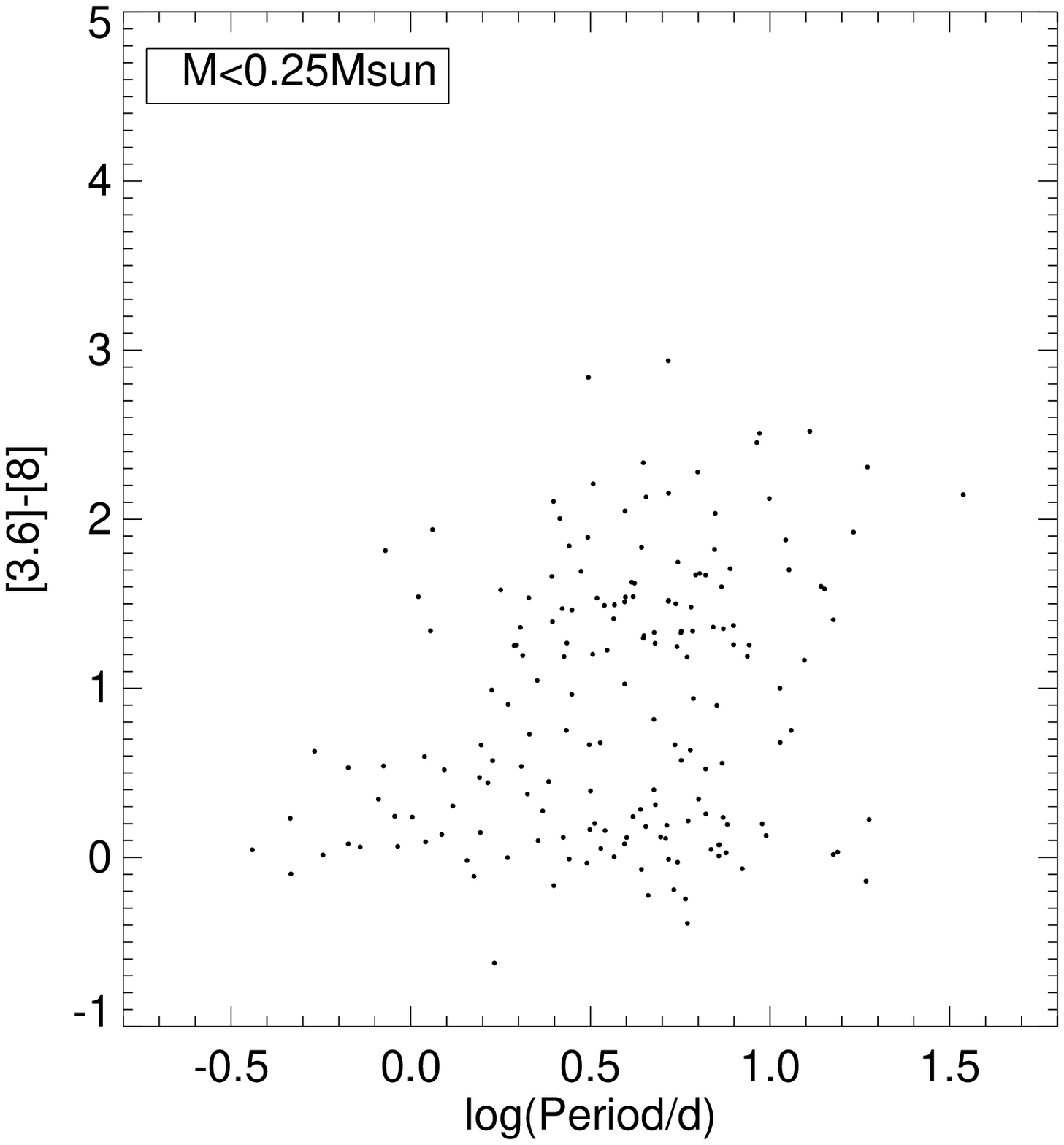}
\caption{Plots similar to Fig.~\ref{fig:p14}, but for stars with
masses $>$ and $<$ 0.25 \msun\ (type M2.5-M3; see text).  The
lower-mass stars have more scatter, and a less well-defined disk
population.  The clump of higher-mass stars near log
$P\sim$0.5-1 may suggest that an average disk ceases to
influence the star at longer periods for higher mass stars. }  
\label{fig:mass}  
\end{figure}

As seen in Figure~\ref{fig:mass}, stars with $M<$0.25 \msun\
have a higher disk fraction, and the separation between the disk
and non-disk candidates, while still readily apparent, is less
sharp; e.g., there are fractionally more lower-mass stars with
$[3.6]-[8]$ values near 1.  There is more scatter in the
lower-mass stars; the errors in periods are comparably small for
all masses, but the errors in IRAC color are likely to be
greater for the fainter stars.  The fastest rotators still tend
to lack disks.  A two-dimensional K-S test indicates that there
is $<$1 chance in $10^5$ that these distributions were drawn
from the same population. There is a well-defined clump of
higher-mass stars with IR excesses between log $P\sim$0.6 and 1,
which does not appear with such clarity in the lower-mass
stars.   The {\em mean} period for the distribution of stars
with disks is identical within the scatter for both mass ranges
(log $P$=0.8$\pm$0.3 for the more massive stars and 0.7$\pm$0.3
for the less massive).  The {\em modal} (or most likely) period
of these distributions of stars with disks are slightly
different; for the more massive stars, the mode of the
distribution is log $P$=0.9 (with 33\% of the sample having log
$P$=0.9$\pm$0.05) and 0.6 for the less massive stars (this
distribution is broader, so although the most frequent value is
0.6, only 15\% of the sample has log $P$=0.6$\pm$0.05).   

The higher mass ``clump'' suggests consistency with Taurus
(Edwards \etal\ 1993) in the sense that the Taurus sample was
dominated by K7-M0 stars, as is the left side of
Fig.~\ref{fig:mass}.  We can speculate on many possible
explanations for this.  It may be that disk locking is more
efficient in the more massive stars, or that the parameters
(\mdot, $B$) relevant for disk locking found in these systems
are more similar across stars in this mass range, such that the
locking occurs at similar radii and periods for a larger
fraction of the sample in the higher mass stars.  

As a function of age within this sample, there is no
distinguishable difference in the clear separation of disk
candidates from non-disk candidates, or in the tendency for fast
rotators to lack disks.  Since younger stars may have a higher
disk fraction, there is the expectation that there may also be
some subtle second-order effects from these influences, but it
is still true, regardless of the subsample, that faster rotators
are less likely to have disks.

\begin{figure}
\plotone{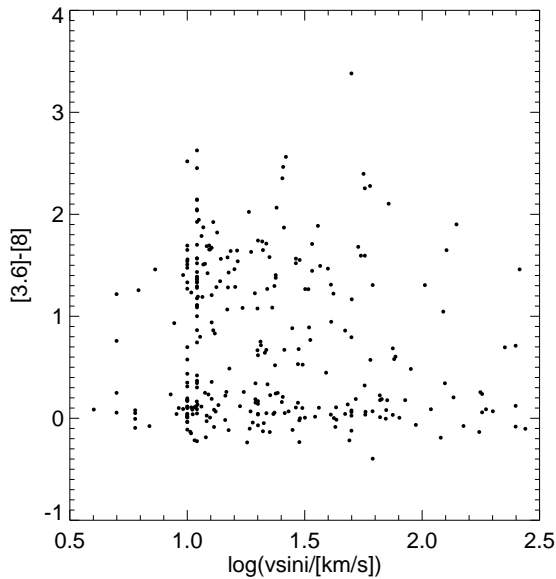}
\caption{Plot similar to Fig.~\ref{fig:p14}, but for stars with
measured \vsini. As before, there is a clear separation of the
disk candidates; there is not as clear a separation between the
fast and slow rotators. This is due primarily to the fundamental
resolution limits imposed by spectroscopy at both the fast- and
slow-rotating ends of the distribution.  Nonetheless, stars that
are more rapidly rotating are still less likely to have disks.
}  
\label{fig:vsini}  
\end{figure}

Similar results can be derived from the independently obtained
distributions of $v \sin i$ in this region; see
Figure~\ref{fig:vsini}.  The relation is not nearly as clear
here, due primarily to the fundamental resolution limits imposed
by spectroscopy at both the fast-rotating (due to smearing out
of the absorption lines in these late-type stars) and
slow-rotating (instrumental resolution is usually 10-12 km
s$^{-1}$) ends of the distribution.  Nonetheless, stars that are
more rapidly rotating are still less likely to have disks.  The
fact that this effect is also seen in the independent \vsini\
sample is powerful confirmation; of the 296 stars with measured
\vsini\ and IRAC fluxes, only 154 of them also have measured
periods.

\section{Discussion \& Conclusions}
\label{sec:disc}

Basic conservation of angular momentum during the star formation
process requires that the angular momentum of the central object
must be regulated.  Recent empirical studies (Rebull \etal\
2004, Herbst \etal\ 2005) have found that young low-mass PMS
objects must be draining angular momentum into their
surroundings, but it has not been clear, until now, that the
disks were the culprit.

The sample discussed here provides two significant advantages
above and beyond previous studies: (1) the disk indicator we use
here is robust against many of the problems plaguing
shorter-wavelength disk indicators, and (2) the sample size is
large, nearly 500 stars.  We have finally resolved two of the
most important issues complicating earlier similar
investigations.  We find a clear correlation between rotation
rate and circumstellar disks.

Among the slower rotators (stars with periods $>$ 1.8d), the
period distributions for stars with and without disks
([3.6]$-$[8]$>$ and $<$1) are statistically indistinguishable. 
Edwards \etal\ (1993) made a similar observation (with many
fewer objects) in their early discussion of disk-locking, and
speculated that the long-period non-disked objects have been
released recently (within a few hundred thousand years). If this
speculation is correct, then examination of the period
distributions among older star-forming regions should reveal a
significant separation in the period distributions betwen disked
and non-disked stars. The evolutionary timescales here may be
important, as suggested by Hartmann (2002); the timescales for
angular momentum loss may be comparable to the evolutionary
and/or disk clearing timescales.  Especially if these stars have
recently cleared their disks, it is possible that they are still
behaving as disk-locked; we could have disk {\em braking}
without disk {\em locking}.  If this is the case, then the
timescales for disk clearing are indeed comparable to the
evolutionary timescales here.  Another possible explanation, as
Long \etal\ (2005) have suggested, is that combinations of
parameters (\mdot, $B$) could result in disk locking at
different rotation rates for otherwise similar stars.  Finally,
it could be important to consider if these stars undergo
intermittent periods of spinup and/or rapid accretion (e.g.,
Lovelace \etal\ 1995, Li 1996, Kuker \etal\ 2003, Long \etal\
2005, Matt \& Pudritz 2005b), not just in terms of whether we
have caught some stars in the act of spinning up, but also
specifically the implications for disk luminosity in the IRAC
bands during periods of high accretion rate.

Given the observed correlation, disks are the most obvious
solution, but it is possible that there are other explanations. 
If the rapid rotators are more likely to be in close binary
systems, then this correlation could be explicable in terms of
disk dissipation by the companion and/or source confusion in
determining the IRAC colors. Followup high-resolution
spectroscopy of the fast rotators looking for indications of
binarity will be required to resolve this issue.

The remaining issues to pursue in similar investigations include
sampling (a) a broader range of environmental settings, e.g.,
regions further from (or lacking entirely) O and B stars, and
(b) a broader range of ages.  If the characteristics (e.g.,
ionization fraction, \mdot, lifetime) of the disks around young
stars are intimately tied to the net time spent near O and B
stars, the distribution found in Orion may be significantly
different from those found in other clusters lacking high-mass
stars. If the observations here in Orion are representative of
the characteristics of other young low-mass systems, we expect
that the difference in period distributions of stars with and
without MIR excesses will increase with time, as suggested by
the Monte Carlo models discussed in Rebull \etal\ (2004).  In
older clusters, there should be a lower fraction of slowly
rotating stars lacking disks, since those stars that may be
``just released'' here should have spun up in older clusters. 
In order to constrain the timescale for disk locking (and
understand whether or not there is an effect with cluster age)
by comparing Orion to other clusters, ideally, we want to
minimize errors by comparing IRAC excess distributions in other
clusters, not those of other disk indicators.  Additional
clusters with substantial numbers of measured periods and
Spitzer observations are NGC 2264 ($\sim$3 Myr), IC 348 ($\sim$3
Myr) and Taurus-Auriga ($\sim$1 Myr).

\begin{acknowledgements}
L.M.R.\ wishes to acknowledge many helpful conversations with S.\
E.\ Strom and S.\ C.\ Wolff.
This work is based in part on observations made with the Spitzer
Space Telescope, which is operated by the Jet Propulsion Laboratory,
California Institute of Technology under a contract with NASA.
Support for this work was provided by NASA through an award issued by
JPL/Caltech.
This research has made extensive use of NASA's Astrophysics Data System
Abstract Service.   
The research described in this paper was partially carried out at the
Jet Propulsion Laboratory, California Institute of Technology, under
a contract with the National Aeronautics and Space Administration.
\end{acknowledgements}

\clearpage

\clearpage

\clearpage

\clearpage

\clearpage

\clearpage

\clearpage

\begin{deluxetable}{rrccccccccc}
\tabletypesize{\scriptsize}
\tablecolumns{7}
\tablewidth{0pt}
\tablecaption{Stars detected in 3.6 and 8 microns with periods
from the literature\tablenotemark{a}
\label{tab:data}}
\tablehead{
\colhead{star\tablenotemark{b}}& \colhead{position (J2000)} & \colhead{[3.6]}& 
\colhead{error} & \colhead{[4.5]}&  
\colhead{error} & \colhead{[5.8]}&  
\colhead{error} & \colhead{[8.0]}& \colhead{error} & \colhead{Period (d)} }
\startdata
  R01- 678 & 05 33 36.9  -05 23 06.2 & 11.52  & 0.006 & 11.52 & 0.008  
  & 11.47 & 0.027 & 11.44 & 0.119 &  7.23\\
\enddata
\tablenotetext{a}{Table will be presented in its entirety in the
electronic version. 
Headings are shown here for guidance regarding its content. }
\tablenotetext{b}{R01 numbers come from Rebull (2001), HBC
numbers from Herbig \& Bell (1988), Par numbers from Parenago
(1954), CHS numbers from Carpenter \etal\ (2001), H97 numbers
from Hillenbrand (1997), HBJM numbers from Herbst \etal\ (2001),
and JW numbers come from Jones \& Walker (1988).}
\end{deluxetable}

\end{document}